\documentclass[preprint,showpacs,preprintnumbers,amsmath,amssymb]{revtex4}

\usepackage{graphicx}
\usepackage{epsfig}		
\usepackage{dcolumn}
\usepackage{bm}
\def\0{\mbox{\tiny $0$}}
\def\1{\mbox{\tiny $1$}}
\def\2{\mbox{\tiny $2$}}
\def\3{\mbox{\tiny $3$}}
\def\4{\mbox{\tiny $4$}}
\def\5{\mbox{\tiny $5$}}
\def\6{\mbox{\tiny $6$}}
\def\7{\mbox{\tiny $7$}}
\def\8{\mbox{\tiny $8$}}
\def\9{\mbox{\tiny $9$}}

\def\n{\mbox{\tiny $n$}}

\def\f14{\mbox{\tiny $\frac{1}{4}$}}

\def\N{\mbox{\tiny $N$}}
\def\U{\mbox{\tiny $U$}}
\def\F{\mbox{\tiny $F$}}
\def\R{\mbox{\tiny $R$}}

\def\s{\mbox{\tiny $s$}}

\def\mi{\mbox{\tiny $-$}}

\def\pl{\mbox{\tiny $+$}}
\def\al{\mbox{\tiny $\alpha$}}

\def\bb#1{\mbox{\footnotesize $(#1)$}}

\begin{document}

\preprint{DF/IST-10.2007}
\preprint{December 2007}

\title{A perturbative approach for mass varying neutrinos coupled to the dark sector in the generalized Chaplygin gas scenario}

\author{A. E. Bernardini}
\affiliation{Departamento de F\'{\i}sica, Universidade Federal de S\~ao Carlos, PO Box 676, 13565-905, S\~ao Carlos, SP, Brasil}
\email{alexeb@ifi.unicamp.br}
\author{O. Bertolami}
\affiliation{Instituto Superior T\'ecnico, Departamento de F\'{\i}sica, Av. Rovisco Pais, 1, 1049-001, Lisboa, Portugal}
\email{orfeu@cosmos.ist.utl.pt}
\altaffiliation[Also at]{~Instituto de Plasmas e Fus\~{a}o Nuclear}

\date{\today}

\begin{abstract}
We suggest a perturbative approach for generic choices for the universe equation of state and introduce a novel framework for studying mass varying neutrinos (MaVaN's) coupled to the dark sector.
For concreteness, we examine the coupling between neutrinos and the underlying scalar field associated with the generalized Chaplygin gas (GCG), a unification model for dark energy and dark matter.
It is shown that the application of a perturbative approach to MaVaN mechanisms translates into a constraint on the coefficient of a linear perturbation, which depends on the ratio between a neutrino energy dependent term and scalar field potential terms.
We quantify the effects on the MaVaN sector by considering neutrino masses generated by the seesaw mechanism.
After setting the GCG parameters in agreement with general cosmological constraints, we find that the squared speed of sound in the neutrino-scalar GCG fluid is naturally positive.
In this scenario, the model stability depends on previously set up parameters associated with the equation of state of the universe.
Our results suggest that the GCG is a particularly suitable candidate for constructing a stable MaVaN scenario.
\end{abstract}

\date{\today}

\maketitle

\normalsize
\section{Introduction}

Substantial observational evidence arising from Type-Ia supernova data \cite{Rie98,Per99,Ast06}, big-bang nucleosynthesis constraints \cite{Bur01}, cosmic microwave background radiation (CMBR) power spectrum \cite{DeB00,Ben03,Spe03}, large scale structure \cite{Pea01}, and determinations of the matter density \cite{Car99,Tur00} suggest a cosmological model where the energy density ($\rho$) of the universe is dominated by an overall smoothly distributed component with negative pressure ($p$), the so-called dark energy, which leads to an accelerated expansion.

At the current stage of our knowledge, speculations about the nature of the dark energy and the negative pressure component responsible for the accelerated expansion of the universe have given origin to a fruitful discussion in the literature \cite{Zla98,Wan99,Ste99,Bar99,Ber00}.
The most obvious explanation for dark energy is the cosmological constant, a constant vacuum energy density which has as equation of state,  $p = -\rho$, at all times.
However, since the cosmological constant has a magnitude completely different from that predicted by theoretical considerations, physicists have been compelled to consider other explanations for the dark sector \cite{Ame02,Kam02,Bil02,Ber02,Cal03,Mot04,Bro06A}.
A plausible alternative for obtaining a negative pressure equation of state, better motivated by the high energy physics, considers a dynamical energy density governed by a light scalar field rolling down in a fairly flat potential \cite{Wet87,Pee87,Rat87}.
These models assume that the vacuum energy can vary \cite{Bro33}.
Guided by theoretical as well as phenomenological arguments, several possibilities have been proposed, such as $k$-essence \cite{Chi00,Arm01}, phantom energy models \cite{Sch01,Car03}, and also several types of modifications of gravity \cite{Def02,Car04,Ama06}.

In this context, one of the most challenging issues concerns models of mass varying neutrinos (MaVaN's) \cite{Hun00,Gu03,Far04,Bja08} coupled to the dark energy light scalar field component.
The simplest realization of the MaVaN mechanism consists in writing down an effective potential which, in addition to a scalar field dependent term, contains a term related to the neutrino energy density.

On the other hand, depending on an eventual compatibility with well-known cosmological bounds on neutrino masses \cite{Goo06,Sko06,Han06,Les06}, for some choices of neutrino-scalar field couplings and scalar field potential, which is intermediated by the so-called {\em stationary condition} \cite{Far04}, the combined fluid (dark energy plus neutrinos) is subject to instabilities once the neutrinos become non-relativistic (NR).
Due to the instability problem \cite{Bja08,Tak06}, models so far proposed usually face fine-tuning problems, which impose severe constraints on the choice of the scalar field potential and their kinetic energy \cite{Pec05}.
Indeed, when a scalar field potential is tuned in agreement with the abovementioned stationary condition, an effective potential, which includes a cosmological term dynamically dependent on the scalar field and a neutrino contribution, is supposed to have a minimum with a steep second derivative for a finite scalar field vacuum expectation value \cite{Bja08}.
If this condition is not satisfied, an attractive force between neutrinos intermediated by the scalar field can lead to the formation of neutrino nuggets \cite{Afs05}.
An immediate consequence is that the coupled fluid would behave as cold dark matter and not as a suitable dark energy candidate.

Notice that dark matter has not been considered in the formulation of the MaVaN mechanisms.
However, the idea of unifying dark energy and dark matter naturally offers this possibility.
The generalized Chaplygin gas (GCG) \cite{Kam02,Bil02,Ber02} is particularly relevant in this respect as it is shown to be consistent with the observational constraints from CMB \cite{Ber03}, supernova \cite{Sup1,Ber04,Ber05}, gravitational lensing surveys \cite{Ber03B}, and gamma ray bursts \cite{Ber06B}.
Moreover, it has been shown that the GCG model can be accommodated within the standard structure formation mechanism \cite{Kam02,Ber02,Ber04}.

In this paper we consider the possibility of neutrino masses arising from an interaction with a real scalar field which describes the dynamics of the generalized Chaplygin gas \cite{Kam02,Ber04}.
As is well known, the Higgs sector \cite{Ber08} and the neutrino sector are possibly the only ones where one can couple a new standard model (SM) singlet without upsetting the known phenomenology.
As it shall be seen, one can treat the mass varying neutrino term in the equation of energy conservation as a perturbative contribution to the evolution of the previously unperturbed adiabatic solution, as for obvious phenomenological reasons, we expect a small contribution from neutrinos to the energy dynamics of the universe.
In particular, we suggest that the dynamics of the scalar field is modified by a linear perturbation over the field itself due to the neutrino mass coupling, which is actually {\em turned on} when the neutrinos become NR.

The basic steps for introducing this perturbative approach as well as the procedure for its implementation is presented in section II.
The applicability of this procedure is, on general terms, compared with the applicability of the stationary condition in the extended Fardon-Nelson-Weiner (FNW) framework \cite{Far04}.
In section III, we review the main features of the GCG, its underlying scalar field ($\phi$), and its dependence on the universe scale factor, $a$.
In section IV, we discuss the neutrino mass generation mechanism in the context of the GCG model.
For this purpose we consider left-handed non-sterile neutrino masses with an analytical dependence on $1/\phi$ and $1/\phi^{\2}$ as suggested by the simplest version of the see-saw mechanism.
In section V, we set the adequate phenomenological constraints for the neutrino masses and for the GCG parameters, consistent with the most recent observational bounds.
Finally, in section VI, we draw our conclusions.

\section{A perturbative approach for cosmological neutrinos}

Cosmological neutrinos have not yet been observed, so hints about their nature and behaviour are quite welcome.
In the usual MaVaN framework, neutrinos are coupled to a light scalar field which is identified with the dark sector.
Presumably, the neutrino mass $m_{\nu}$ has its origin on the vacuum expectation value (VEV) of the scalar field and, naturally, its behaviour is governed by the dependence of the scalar field on the scale factor.
Since the neutrino statistical distribution corresponds to a Fermi-Dirac statistics without a chemical potential $f\bb{q}$, where $q \equiv \frac{|\mbox{\boldmath$p$}|}{T_{\nu\0}}$, $T_{\nu\0}$ being the neutrino background temperature at present, the neutrino energy density and pressure can be expressed in the following way
\begin{eqnarray}
\rho_{\nu}\bb{a, \phi} &=&\frac{T^{\4}_{\nu \0}}{\pi^{\2}\,a^{\4}}
\int_{_{0}}^{^{\infty}}{\hspace{-0.3cm}dq\,q^{\2}\,\left(q^{\2}+\frac{m^{\2}\bb{\phi}\,a^{\2}}{T^{\2}_{\nu\0}}\right)^{\1/\2}\hspace{-0.1cm}f\bb{q}},\\
p_{\nu}\bb{a, \phi} &=&\frac{T^{\4}_{\nu \0}}{3\pi^{\2}\,a^{\4}}\int_{_{0}}^{^{\infty}}{\hspace{-0.3cm}dq\,q^{\4}\,\left(q^{\2}+\frac{m^{\2}\bb{\phi}\,a^{\2}}{T^{\2}_{\nu \0}}\right)^{\mi\1/\2}\hspace{-0.1cm} f\bb{q}},~~~~ \nonumber
\label{gcg01}
\end{eqnarray}
where we have assumed a flat FRW cosmology and introduced the sub-index $0$ for denoting present-day values, with $a_{\0} = 1$.
For simplicity, we have considered a single non-vanishing neutrino mass, although the generalization to more than one massive neutrino is straightforward.

Simple mathematical manipulation allows one to easily demonstrate that
\begin{equation}
m_{\nu}\bb{\phi} \frac{\partial \rho_{\nu}\bb{a, \phi}}{\partial m_{\nu}\bb{\phi}} = (\rho_{\nu}\bb{a, \phi} - 3 p_{\nu}\bb{a, \phi}),
\label{gcg02}
\end{equation}
where, for the neutrino NR regime,
\begin{equation}
\frac{\partial \rho_{\nu}\bb{a, \phi}}{\partial m_{\nu}\bb{\phi}}\simeq n_{\nu}\bb{a} \propto{a^{\mi\3}}.
\label{gcg02B}
\end{equation}
From the dependence of $\rho_{\nu}$ on $a$, one can obtain the energy-momentum conservation for the neutrino fluid,
\begin{equation}
\dot{\rho}_{\nu}\bb{a, \phi} + 3 H (\rho_{\nu}\bb{a, \phi} + p_{\nu}\bb{a, \phi}) =
\dot{\phi}\frac{d m_{\nu}\bb{\phi}}{d \phi} \frac{\partial \rho_{\nu}\bb{a, \phi}}{\partial m_{\nu}\bb{\phi}},
\label{gcg03}
\end{equation}
where $H = \dot{a}/{a}$ is the expansion rate of the universe and the {\em overdot} denotes differentiation with respect to time ($^{\cdot}\, \equiv\, d/dt$).

It is important to emphasize that the coupling between cosmological neutrinos and the scalar field as described by Eq.~(\ref{gcg02}) is restricted to times when neutrinos are NR, i. e. $\frac{\partial \rho_{\nu}\bb{a, \phi}}{\partial m_{\nu}\bb{\phi}} \simeq n_{\nu}\bb{a} \propto{a^{\mi\3}}$ \cite{Far04,Bja08,Pec05}.
In opposition, as long as neutrinos are relativistic ($T_{\nu}\bb{a} = T_{\nu \0}/a >> m_{\nu}\bb{\phi\bb{a}}$), the decoupled fluids should evolve adiabatically since the strength of the coupling is suppressed by the relativistic increase of pressure ($\rho_{\nu}\sim 3 p_{\nu}$).
In this case, one would have
\begin{equation}
\dot{\rho}_{\nu, \phi} + 3 H (\rho_{\nu, \phi} + p_{\nu, \phi}) = 0,
\label{gcg04}
\end{equation}
where for the scalar field background fluid,
\begin{eqnarray}
\rho_{\phi} &=& \frac{\dot{\phi}^{\2}}{2} + V\bb{\phi},\nonumber\\
p_{\phi} &=& \frac{\dot{\phi}^{\2}}{2} - V\bb{\phi}.
\label{gcg05}
\end{eqnarray}

Treating the system of NR neutrinos and the scalar field as a unified fluid (UF) adiabatically expanding with energy density $\rho_{\U\F} = \rho_{\nu} + \rho_{\phi}$ and pressure $p_{\U\F} = p_{\nu} + p_{\phi}$ lead to
\begin{equation}
\dot{\rho}_{\U\F} + 3 H (\rho_{\U\F} + p_{\U\F}) = 0~~ \Rightarrow ~~\dot{\rho}_{\phi} + 3 H (\rho_{\phi} + p_{\phi}) = -\dot{\phi}\frac{d m_{\nu}}{d \phi} \frac{\partial \rho_{\nu}}{\partial m_{\nu}},
\label{gcg06}
\end{equation}
where the last step is derived from the substitution of Eq.~(\ref{gcg03}) into the first equation in (\ref{gcg06}).

The existence of a scalar field dark energy sector on its own constitutes a problem in what concerns order of magnitude equality with the energy densities of the other components of the universe.
The theoretical assumptions proposed in Ref. \cite{Far04} and subsequently developed by other authors \cite{Pec05,Bro06A,Bja08,Tak06} suggest a stationary condition which allows circumventing the coincidence problem for cosmological neutrinos in a way that the dark energy is always diluted at the same rate as the neutrino fluid, that is,
\begin{equation}
\frac{d V\bb{\phi}}{d {\phi}} = - \frac{d m_{\nu}}{d \phi} \frac{\partial \rho_{\nu}}{\partial m_{\nu}}.
\label{gcg07}
\end{equation}
This introduces a constraint over the neutrino mass since it promotes it into a dynamical quantity, as indicated in Eq.~(\ref{gcg06}).
The main feature of this scenario \cite{Far04} is that it is equivalent as adopting cosmological constant like dark sector, with an energy density that varies as a function of the neutrino mass.
As already mentioned, the effectiveness of this coupling is restricted to values of the scale factor larger than $a_{\N\R}$, where $a_{\N\R}$ parameterizes the transition between the relativistic and NR regimes.

The assumption of a universe with the dark energy sector governed by the equation of state $p_{\phi} = -\rho_{\phi}$ implies, through Eq.~(\ref{gcg07}), that  $\rho_{\Lambda} = V$, and allows to recover the stationary condition, exactly as obtained in Ref. \cite{Far04}.
It is clear that the relevance and the considerations about the constraints on the equation of state are actually dominated by the competition among scalar field potentials and its adequacy for neutrino mass generation.
In fact, once one assumes that $p_{\phi} = -\rho_{\phi}$, the neutrino mass evolution and the form of the potential become automatically entangled by the stationary condition.
Thus, it should be realized that the stationary constraint Eq.~(\ref{gcg07}) is quite dependent on the potential of the scalar field.

Moreover, the introduction of a kinetic energy component modifies the equation of state and implies that the stationary condition is not satisfied when $p_{\phi} + \rho_{\phi} = \dot{\phi}^{\2}$ is non-vanishing.
This difficulty has already been pointed out in Ref. \cite{Pec05}, where a solution which severally constrains the choice of the scalar field potential is proposed.
It is then shown that the FNW scenario is consistent only for a vanishing kinetic energy contribution, for a dark energy fluid behaving like a ``running cosmological constant''.

It follows from this discussion that, at least from a theoretical point of view, an alternative approach to treat deviations from the FNW proposal is needed to overcome the quite restrictive condition $\rho_{\phi} + p_{\phi} = 0$, which, in turn, implies in a vanishing kinetic energy contribution.
In spite of being, at present, negligible in comparison to the potential terms, this contribution can dominate and affect the evolution of the universe at earlier times.
Scalar field dark sector candidates with a well defined equation of state, such as for instance the GCG, are natural alternatives to circumvent this problem.
Candidate models can be easily implemented by means of an assumption that the neutrino coupling to the underlying scalar field of the dark energy sector is perturbative.
In this instance, we start instead of Eq.~(\ref{gcg06}), from the unperturbed equation (\ref{gcg04}), and establish the conditions for treating the neutrino coupling in a perturbative way.
To exemplify this point, let us consider the unperturbed equation of motion for the scalar field,
\begin{equation}
\ddot{\phi} + 3 H \dot{\phi} + \frac{d V\bb{\phi}}{d {\phi}} = 0,
\label{gcg08}
\end{equation}
and assume that the effect of the coupling of the neutrino fluid to the scalar field fluid is quantified by a linear perturbation $\epsilon \phi$ ($|\epsilon| << 1$) such that
\begin{equation}
\phi \rightarrow \varphi \simeq (1 + \epsilon) \phi.
\label{gcg09}
\end{equation}
It then follows the novel equation for the energy conservation
\begin{equation}
\ddot{\varphi} + 3 H \dot{\varphi} + \frac{d V\bb{\varphi}}{d {\varphi}} = -\frac{d m_{\nu}}{d \varphi} \frac{\partial \rho_{\nu}}{\partial m_{\nu}}.
\label{gcg10}
\end{equation}
The explicit dependence of $\varphi$ on $\phi$ is easy to quantify.
Indeed, after a simple manipulation one finds
\begin{eqnarray}
\frac{d V\bb{\varphi}}{d {\varphi}}
&\simeq& (1 + \epsilon)^{\mi \1} \frac{d V\bb{\varphi}}{d {\phi}}\nonumber\\
&\simeq& (1 + \epsilon)^{\mi \1} \frac{d}{d {\phi}} \left[V\bb{\phi} + (\epsilon \phi)\frac{d V\bb{\phi}}{d {\phi}} \right]\nonumber\\
&\simeq& \frac{d V\bb{\phi}}{d {\phi}} + (\epsilon \phi) \frac{d^{\2} V\bb{\phi}}{d {\phi^{\2}}}.
\label{gcg11}
\end{eqnarray}
The substitution of the Eqs.~(\ref{gcg09}) and (\ref{gcg11}) into Eq.~(\ref{gcg10}) and use of Eq.~(\ref{gcg08}), lead to
\begin{eqnarray}
\epsilon \left[\phi \frac{d^{\2} V\bb{\phi}}{d {\phi^{\2}}} - \frac{d V\bb{\phi}}{d {\phi}} \right] \simeq \frac{d m_{\nu}}{d \varphi} \frac{\partial \rho_{\nu}}{\partial m_{\nu}} \simeq \frac{d m_{\nu}}{d \phi}
n_{\nu}\bb{a}
\label{gcg12}
\end{eqnarray}
where the perturbative character of the neutrino mass term is assumed when we set the last approximation in the above equation.
Finally, we can obtain for the value of the coefficient of the perturbation
\begin{eqnarray}
\epsilon  \simeq \frac{-\frac{d m_{\nu}}{d \phi}\frac{\partial \rho_{\nu}}{\partial m_{\nu}}}{\left[\phi^{\2} \frac{d}{d {\phi}}\left(\frac{1}{\phi} \frac{d V\bb{\phi}}{d {\phi}}\right)\right]},
\label{gcg13}
\end{eqnarray}
which for consistency is required to satisfy the condition $|\epsilon| << 1$.
Of course, this procedure can be carried out to higher order as to determine the additional perturbative modifications to $\rho_{\U\F}$.

It is important to notice that in the conventional quintessence models the scalar field is, at the present epoch,
slowly rolling dawn its potential and therefore its effective mass, $(d^{\2}V/d\phi^{\2})^{\1/\2}$ is smaller than the Hubble expansion rate, $H$.
In opposition, the usual SC framework for MaVaN's establishes that the dynamical scalar field sits at the instantaneous minimum of its potential, and the cosmic expansion only modulates this minimum through changes in the local neutrino density \cite{Afs05}.
It hence allows for $(d^{\2}V/d\phi^{\2})^{\1/\2} >> H$, which means that the coherence length of the dynamical scalar field is much smaller than the present Hubble length, and thus, unlike in quintessence models, the perturbations on sub-Hubble scales are adiabatic.
Consequently, the speed of sound, for these perturbations is simply given by $c_{s}^{\2} = d p_{\U\F}/ d \rho_{\U\F}$.
The very argument of a scalar potential being modulated by a local neutrino energy density applies to our approach.
 This is reinforced by the idea that the Higgs \cite{Ber08} and the neutrino sectors are the only ones where one can couple a new standard model (SM) singlet without upsetting the known phenomenology.

Otherwise, our setup allows one to consider a large class of scalar field potentials and equations of state for the dark sector, for which the stationary condition is incompatible with generic neutrino mass generation models.
Furthermore, given its perturbative nature, our procedure is valid for any equation of state satisfying the relevant phenomenological requirements.
In the following analysis one can verify the stability condition by observing that the square of the speed of sound of the coupled fluid is also dominated by the unperturbed equation of state.

In fact, this feature and the ensued problem could have already been observed in the context of the FNW scenario since the stationary condition implies that $c_{s}^{\2} = -1$ and the burden of recovering the stability ($c^2_{s (\nu + \phi)} > 0$) is transferred to the neutrino contribution.

It is easy to see that our result reduces to the FNW scenario when $\dot{\phi} = \dot{\varphi} = 0$ or when one assumes that $\rho_{\phi} = - p_{\phi} = V$.
Indeed, rewriting the equation for the conservation of energy, Eq.~(\ref{gcg06}), by simply redefining $\rho_{\U\F}$ as
\begin{equation}
\rho_{\U\F}  = \frac{1}{2}\dot{\phi}^{\2} + V_{\mbox{\tiny Eff}},
\label{gcg14}
\end{equation}
the usual definition \cite{Bro06A,Far04,Bja08} for an {\em effective} potential $V_{\mbox{\tiny Eff}}$ in terms of $\frac{d V_{\mbox{\tiny Eff}}}{d \phi} = \frac{d V\bb{\phi}}{d \phi} + \frac{d m_{\nu}}{d \phi} \frac{\partial\rho_{\nu}}{\partial m_{\nu}}$ is now valid for any value of $\dot{\phi}$.

In the next section we shall implement this perturbative approach and verify the applicability criteria for the coupling between variable mass neutrinos and the scalar field associated with the GCG background fluid.

\section{The generalized Chaplygin gas - An state equation for the dark sector}

The GCG model is characterized by an exotic equation of state \cite{Ber02,Ber03} given by,
\begin{equation}
p = - A_{\s} \rho_{\0} \left(\frac{\rho_{\0}}{\rho}\right)^{\al},
\label{gcg20}
\end{equation}
which can be obtained from a generalized Born-Infeld action \cite{Ber02}.
In any case, irrespective of its origin, several studies yield convincing evidence that the GCG scenario is a phenomenologically consistent approach to explain the accelerated expansion of the universe.
The constants $A_{\s}$ and $\alpha$ are positive and $0 < \alpha \leq 1$.
Of course, $\alpha = 0$ corresponds to the $\Lambda$CDM model and we are assuming that the GCG model has an underlying scalar field, actually real \cite{Kam02,Ber04} or complex \cite{Bil02,Ber02}.
The case $\alpha = 1$ corresponds to the equation of state of the Chaplygin gas scenario \cite{Kam02} and is already ruled out by data \cite{Ber03}.
Notice that for $A_s =0$, GCG behaves always as matter whereas for $A_{\s} =1$, it behaves always as a cosmological constant.
Hence to use it as a unified candidate for dark matter and dark energy one has to exclude these two possibilities so that $A_s$ must lie in the range $0 < A_{\s} < 1$.

Inserting the above equation of state into the unperturbed energy conservation Eq.~(\ref{gcg04}), one obtains through a straightforward integration \cite{Ber02}
\begin{equation}
\rho_{\phi} = \rho_{\0} \left[A_{\s} + \frac{(1-A_{\s})}{a^{\3(\1+\alpha)}}\right]^{\1/(\1 \pl \al)},
\label{gcg21}
\end{equation}
and
\begin{equation}
p_{\phi} = - A_{\s} \rho_{\0} \left[A_{\s} + \frac{(1-A_{\s})}{a^{\3(\1+\alpha)}}\right]^{-\al/(\1 \pl \al)}.
\label{gcg22}
\end{equation}
Following Ref. \cite{Ber04}, one can obtain through Eq.~(\ref{gcg05}) the field time-dependence,
\begin{equation}
\dot{\phi}^{\2}\bb{a} = \frac{\rho_{\0}(1 - A_{\s})}{a^{\3(\al + \1)}}
\left[A_{\s} + \frac{(1-A_{\s})}{a^{\3(\1+\alpha)}}\right]^{-\al/(\1 \pl \al)},
\label{gcg23}
\end{equation}
and assuming a flat evolving universe described by the Friedmann equation $H^{\2} = \rho_{\phi}$ (with $H$ in units of $H_{\0}$ and $\rho_{\phi}$ in units of $\rho_{\mbox{\tiny Crit}} = 3 H^{\2}_{\0}/ 8 \pi G)$, one obtains
\begin{equation}
\phi\bb{a} = - \frac{1}{2 \beta}\ln{\left[\frac{\sqrt{1 - A_{\s}(1 - a^{\2\beta})} - \sqrt{1 - A_{\s}}}{\sqrt{1 - A_{\s}(1 - a^{\2\beta})} + \sqrt{1 - A_{\s}}}\right]},
\label{gcg24}
\end{equation}
where one assumes that
\begin{equation}
\phi_{\0} = \phi\bb{a_{\0} = 1} = - \frac{1}{2 \beta}\ln{\left[\frac{1 - \sqrt{1 - A_{\s}}}{1 + \sqrt{1 - A_{\s}}}\right]}
\label{gcg25}
\end{equation}
and $\beta = 3(\alpha + 1)/2$.

One then readily find the scalar field potential in terms of the field,
\begin{equation}
V\bb{\phi} = \frac{1}{2}A_{\s}^{\frac{\1}{\1 \pl \al}}\rho_{\0}\left\{
\left[\cosh{\left(\beta \phi\right)}\right]^{\frac{\2}{\al \pl \1}}
+
\left[\cosh{\left(\beta \phi\right)}\right]^{-\frac{\2\al}{\al \pl \1}}
\right\}.
\label{gcg26}
\end{equation}

\begin{figure}
\centerline{\psfig{file=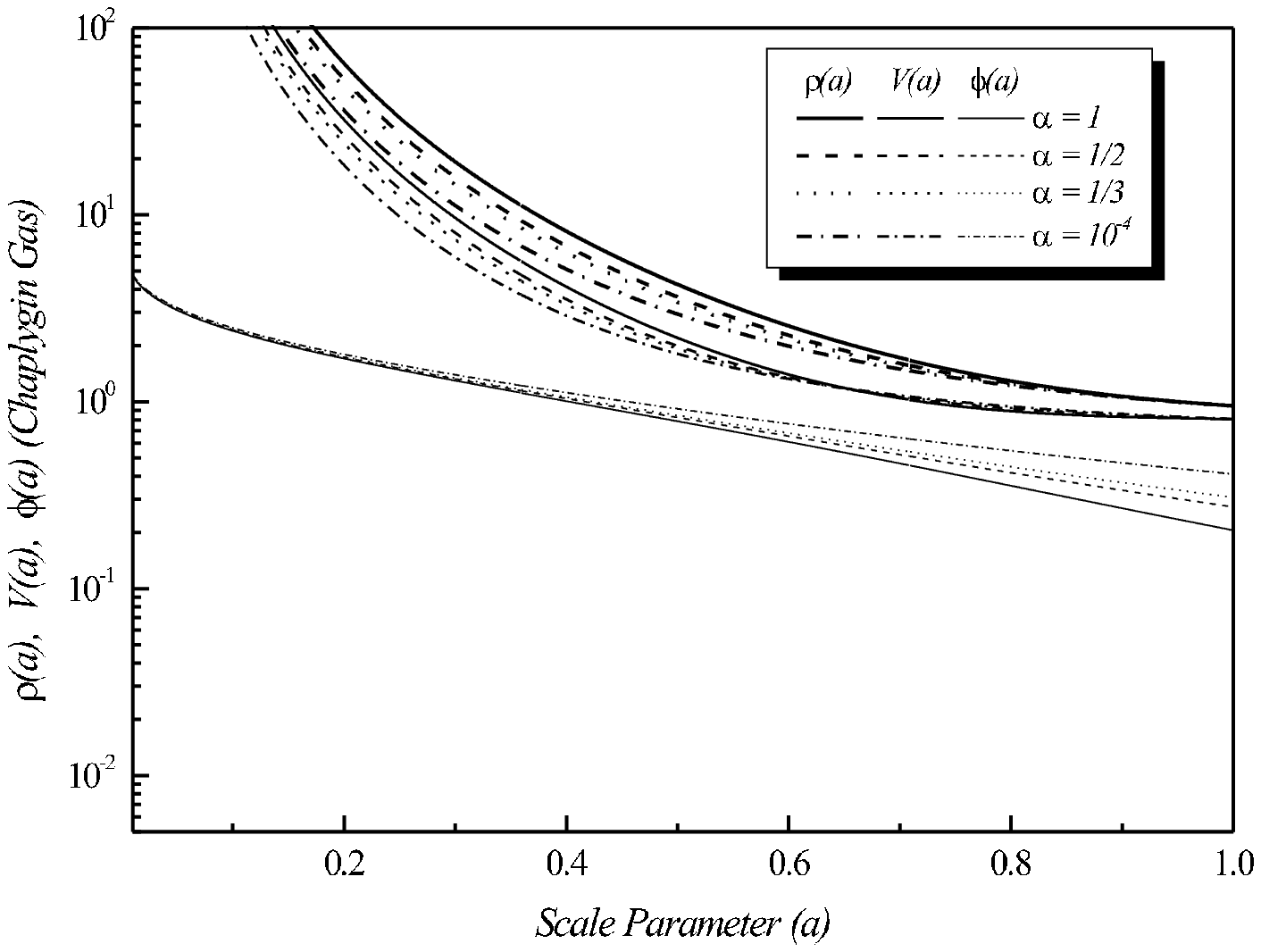,width=14cm}}
\caption{\small
The energy density $\rho_{\phi}\bb{a}$, the potential energy $V\bb{\phi}$ and the scalar field $\phi\bb{a}$ as functions of the scale factor $a$ for several choices of $\alpha$ and for $A_{\s} = 0.7$.}
\label{fGCG-01}
\end{figure}

The analytical dependence of the energy density, the potential energy and the scalar field in terms of the scale factor is illustrated in the Fig.~\ref{fGCG-01} for some particular choices of $\alpha$ and for $A_{\s} = 0.7$.
The latter value arises from the matching of the model with supernova, CMB data, and cosmic topology \cite{Ber03,Ber04,Ber06}.
Notice that in what concerns $\alpha$, the observational constraints are as follows:
WMAP1 is compatible with $\alpha \lesssim 0.6$ \cite{Ber03};
WMAP3 admits values in the range $\alpha \lesssim 0.2$ \cite{Ber07B}; and
structure formation implies that $\alpha \lesssim 0.2$ \cite{Ber04B}.

One of the most striking features of the GCG fluid is that its energy density interpolates between a dust dominated phase, $\rho_{ch} \propto a^{-\3}$, in the past, and a de-Sitter phase, $\rho_{ch} = -p_{ch}$, at late times.
This property makes the GCG model an interesting candidate for the unification of dark matter and dark energy.
Indeed, it can be shown that the GCG model admits inhomogeneities and that, in particular, in the context of the Zeldovich approximation, these evolve in a qualitatively similar way as in the $\Lambda$CDM model \cite{Ber02}.
Furthermore, this evolution is controlled by the model parameters, $\alpha$ and $A_{\s}$.

In order to understand the possible range of values for $\alpha$, one has to consider the propagation of sound through the GCG fluid, $c_{s}^{\2} = d p_{\phi}/ d \rho_{\phi}$.
A detailed quantitative analysis of the stability conditions for the GCG in terms of the squared speed of sound is discussed in Ref. \cite{Ber04}.
Positive $c_{s}^{\2}$ implies that $0 \leq \alpha \leq 1$.
At a later stage, we shall come back to this issue when considering the coupling with neutrinos.

\section{Neutrino mass models}

Despite its impressive phenomenological success, it is widely believed that the SM of particle physics is actually only a low-energy approximation of an underlying more fundamental theory.
In this respect, the interplay with the cosmology can be an important guideline to obtain insights on the nature of the more fundamental theory.
In the SM, the most natural way to explain the smallness of the neutrino masses is through the seesaw mechanism, according to which, the tiny masses, $m_{\nu}$, of the usual left-handed neutrinos are obtained via a very massive, $M$, {\em sterile} right-handed neutrino.
The Lagrangian density that describes the simplest version of the seesaw mechanism through the Yukawa coupling between a light scalar field and a single neutrino flavour is given by
\begin{equation}
{\cal L} = m_{LR} \bar{\nu}_L \nu_R + M\bb{\phi} \bar{\nu}_R \nu_R + h.c.,
\label{gcg30}
\end{equation}
where it is shown that at scales well below the right-handed neutrino mass, one has the effective Lagrangian density \cite{seesaw}
\begin{equation}
{\cal L} = \frac{m_{LR}^{\2}}{M\bb{\phi}} \bar{\nu}_L \nu_R + h.c..
\label{gcg31}
\end{equation}
Phenomenological consistency with the SM implies that logarithm corrections to the above terms are small, while
it is well-know from the results of solar, atmospheric, reactor and accelerator neutrino oscillation experiments that neutrino masses given by $m_{\nu} \sim m_{LR}^{\2}/M\bb{\phi}$ lie in the sub-$eV$ range \cite{neutrino}.
It is also clear that promoting the scalar field $\phi$ into a dynamical quantity, $\phi\rightarrow \varphi\bb{a}$ leads to a mechanism in the context of which neutrino masses are time-dependent.
Associating the scalar field to the dark energy field allows linking NR neutrino energy densities to late cosmological times \cite{Far04,Bro06A,Bja08}.
This scenario can be implemented through the perturbative approach via Eq.~(\ref{gcg06}).
It is evident that this approach is fairly general, as well as independent of the choice of the equation of state and on the dependence of the neutrino mass on the scalar field.
For sure, the form of $M\bb{\phi}$ and of the equation of state can lead to quite different scenarios.
In studying the neutrino coupling with the GCG fluid, we consider two cases, namely, $m_{\nu} = m_{\0} (\phi_{\0}/\phi)$ and $m_{\nu} = m_{\0} (\phi_{\0}/\phi)^{\2}$.
These two scenarios are by no means the only possibilities.
In particular, we choose them as they correspond to the simplest feasible possibility for sterile neutrino effective mass generation, given that the scalar field has mass dimension one \cite{seesaw}.

In particular, the essential information content of the mass dependence on the scale factor is obvious from the fact that neutrino masses increases with decreasing $\phi$, which in the GCG model is a decreasing function of $a$.
In the phenomenological analysis performed in the next section we shall elucidate this point by noticing that for larger values of the scale factor (decreasing red-shift) the neutrino mass increases.

\begin{figure}
\vspace{-0.8 cm}
\centerline{\psfig{file=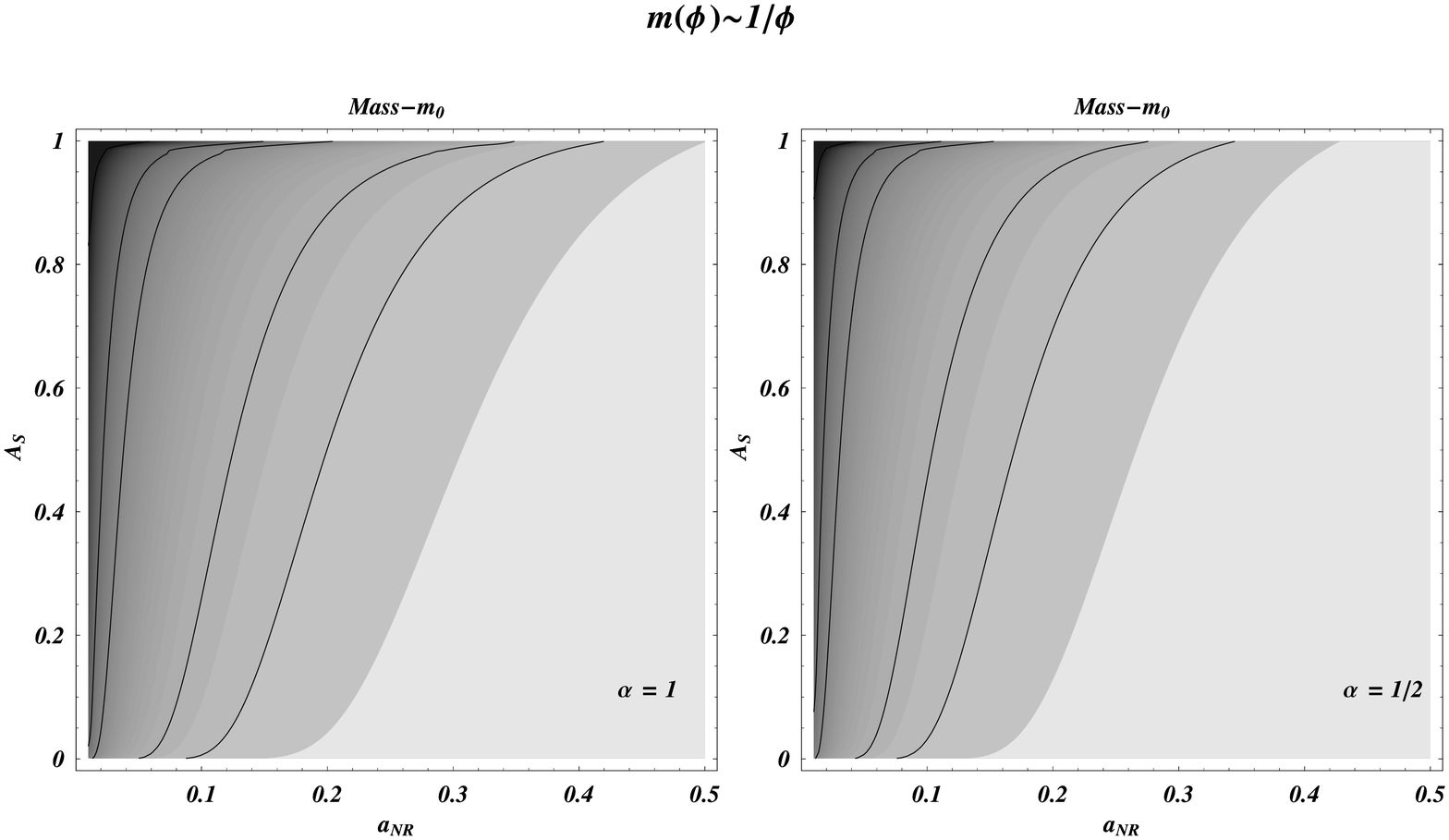,width=13 cm}}
\vspace{-0.5 cm}
\centerline{\psfig{file=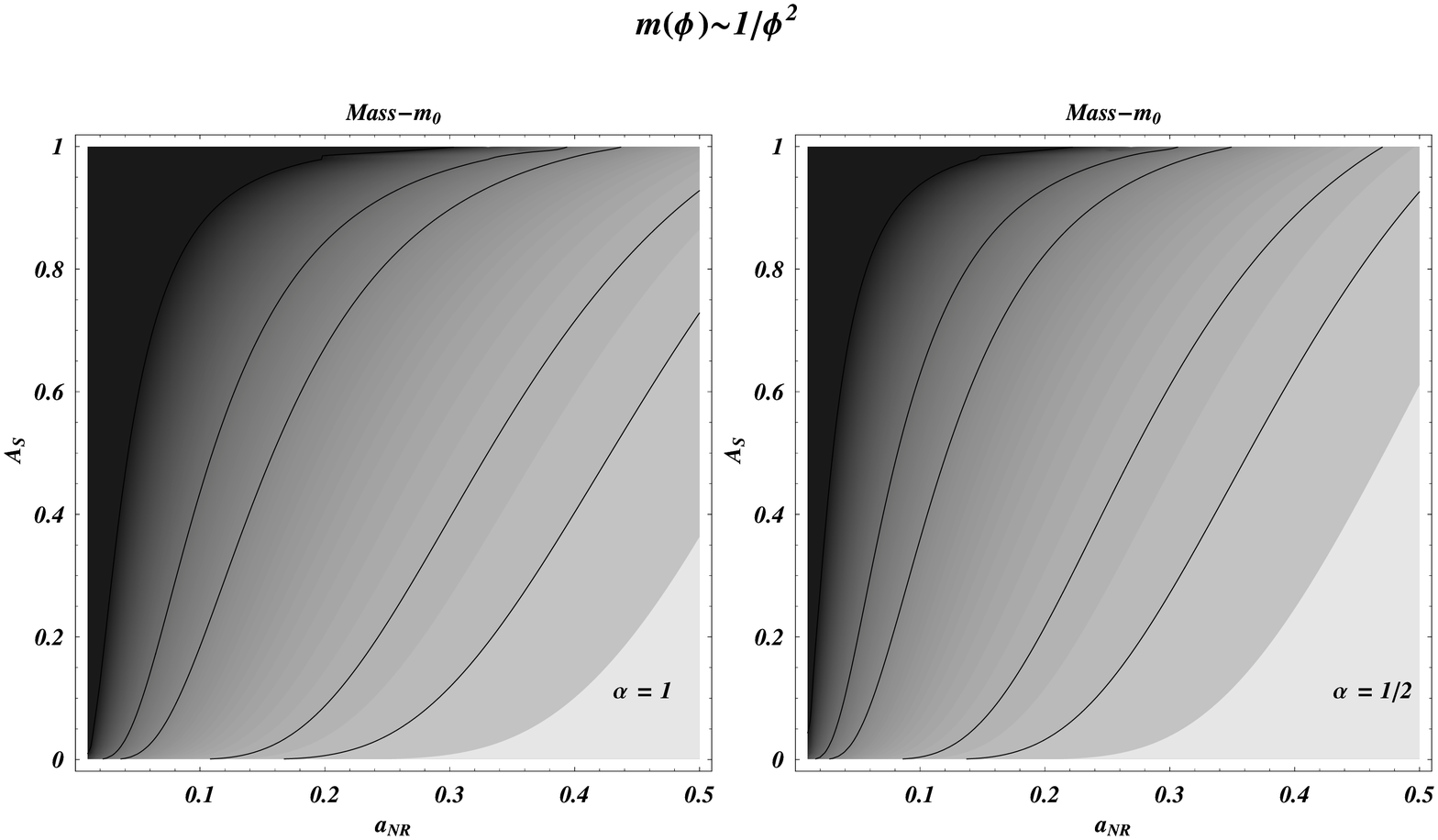,width=13 cm}}
\vspace{-1.0 cm}
\caption{\small Present values of the neutrino mass $m_{\0}$ and the corresponding values of $a_{\N\R}$ for which the transition between the NR and UR regimes takes place in the GCG phenomenological scenario with variable $A_{\s}$ and $\alpha = 1,\,1/2$.
The neutrino mass varies with $1/\phi$ and $1/\phi^{\2}$
and the increasing {\em gray level} corresponds to increasing values of $m_{\0}$, for which the boundary values are $m_{\0} = 0.05\, eV,\,0.1\, eV,\,0.5\, eV,\,1\,eV,\, 5\, eV$.}
\label{fGCG-03}
\end{figure}

In addition to the mass dependence, it is necessary to determine for which values of the scale factor the neutrino-scalar field coupling becomes important.
For convenience we set the value of $a = a_{\N\R}$ for which $\rho_{\nu,\N\R} = \rho_{\nu,\U\R}$ (which sometimes is expressed in terms of $m_{\nu}\sim T_{\nu}$), to parameterize the transition between the NR and the ultra-relativistic (UR) regime.
In fact, this occurs when
\begin{equation}
m_{\nu}\bb{a} = m_{\0}(\phi_{\0}/\phi\bb{a})^{n} = \chi \frac{T_{\nu, \0}}{a}
\label{gcg33}
\end{equation}
where $n = 1,\, 2$ and $\chi$ is a numerical factor estimated to be about $\chi \simeq 94$ supposing that $\rho_{\nu}/\rho_{\mbox{\tiny Crit}} = m_{\0}\,[eV]/(94\,h^{\2}\,[eV])$, where $h$ is the value of the Hubble constant in terms of $100\, km \,s^{\mi \1} Mpc^{\mi \1}$.
This condition allows one to establish the correspondence between the values of $a_{\N\R}$ for which the (UR to NR) transition takes place and the neutrino masses assume the present-day values.
For our model such a transition is shown in Fig.~\ref{fGCG-03} for the studied set of parameters of the GCG model.

\begin{figure}
\vspace{-0.8 cm}
\centerline{\psfig{file=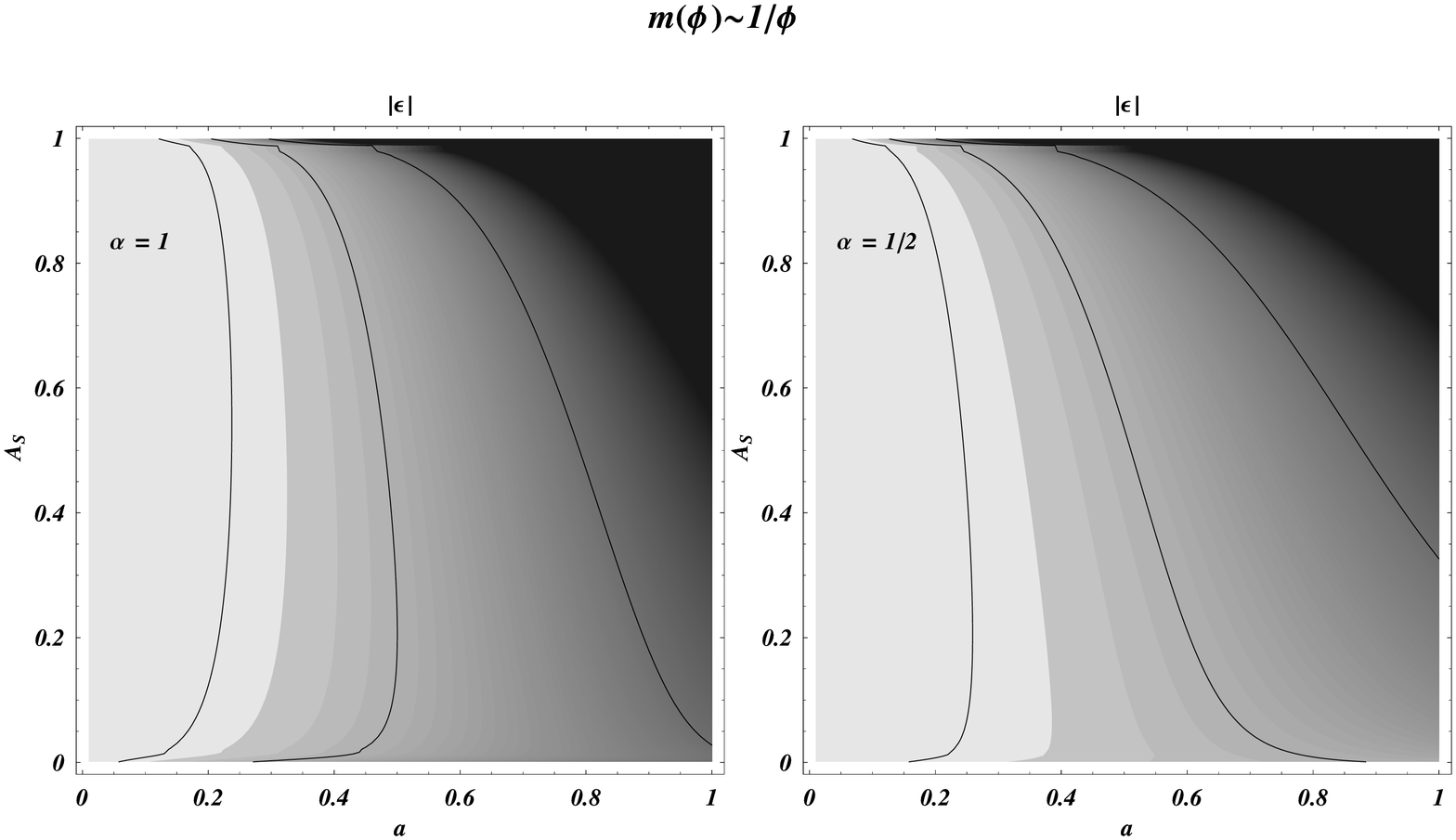,width=13 cm}}
\vspace{-0.5 cm}
\centerline{\psfig{file=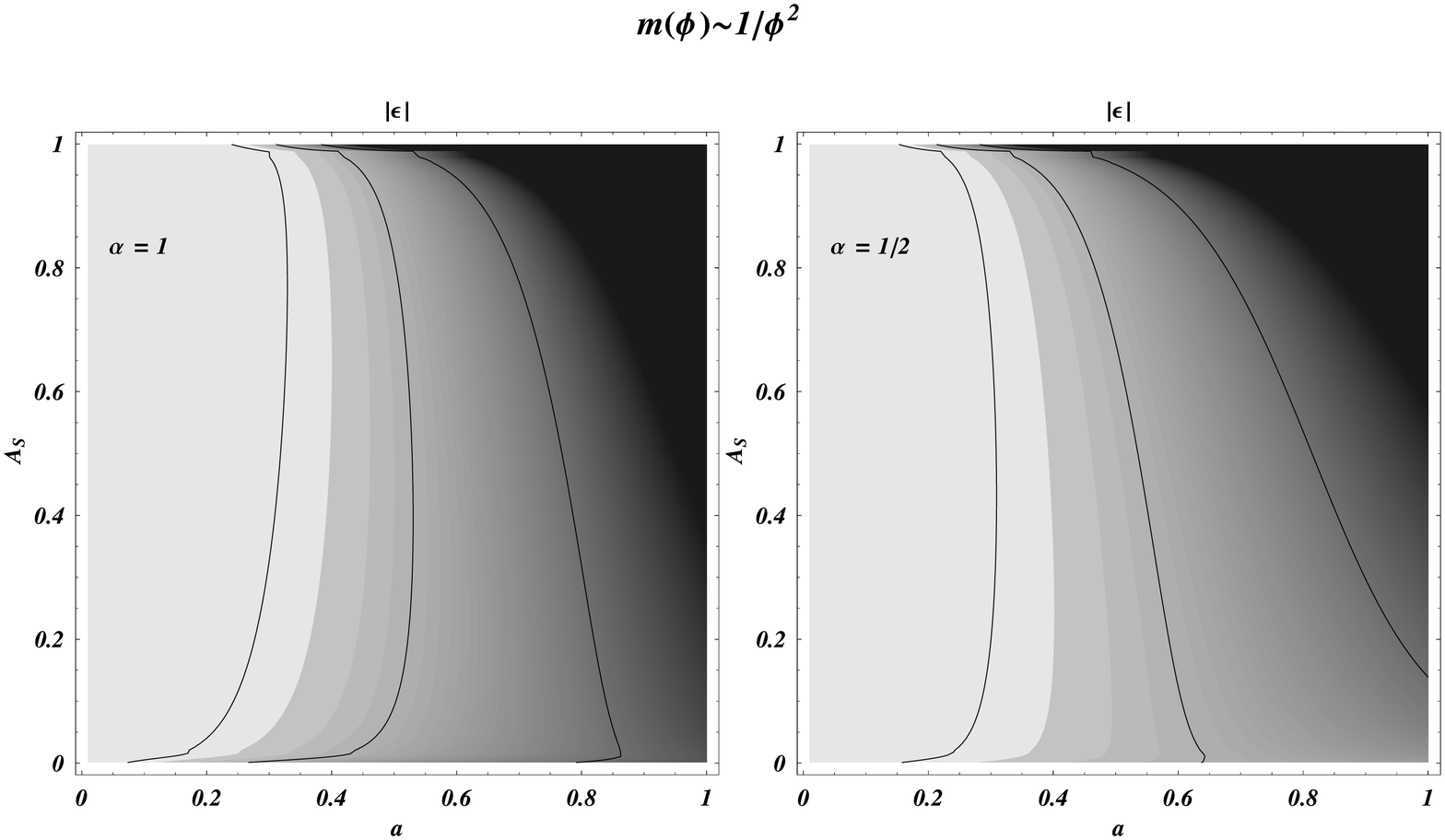,width=13 cm}}
\vspace{-1.0 cm}
\caption{\small Determination of the maximal value of the $|\epsilon|$ parameter as a function of the scale
factor and of the GCG parameters $A_{\s}$ and $\alpha$.
For illustration have set $m_{\0} = 1\, eV$.
The increasing {\em gray level} corresponds to growing values of $|\epsilon|$,
for which we have marked the boundary values $|\epsilon|= 1,\, 0.1,\, 0.01$.
Independently of the neutrino mass dependence on $\phi$ ($m \sim 1/\phi$ or $m\sim 1/\phi^{\2}$), it is clear from the graphs that $\epsilon_{\mbox{\tiny Max}} = \epsilon\bb{a = 1}$ for all that cases we have considered.}
\label{fGCG-04}
\end{figure}

\begin{figure}
\vspace{-0.8 cm}
\centerline{\psfig{file=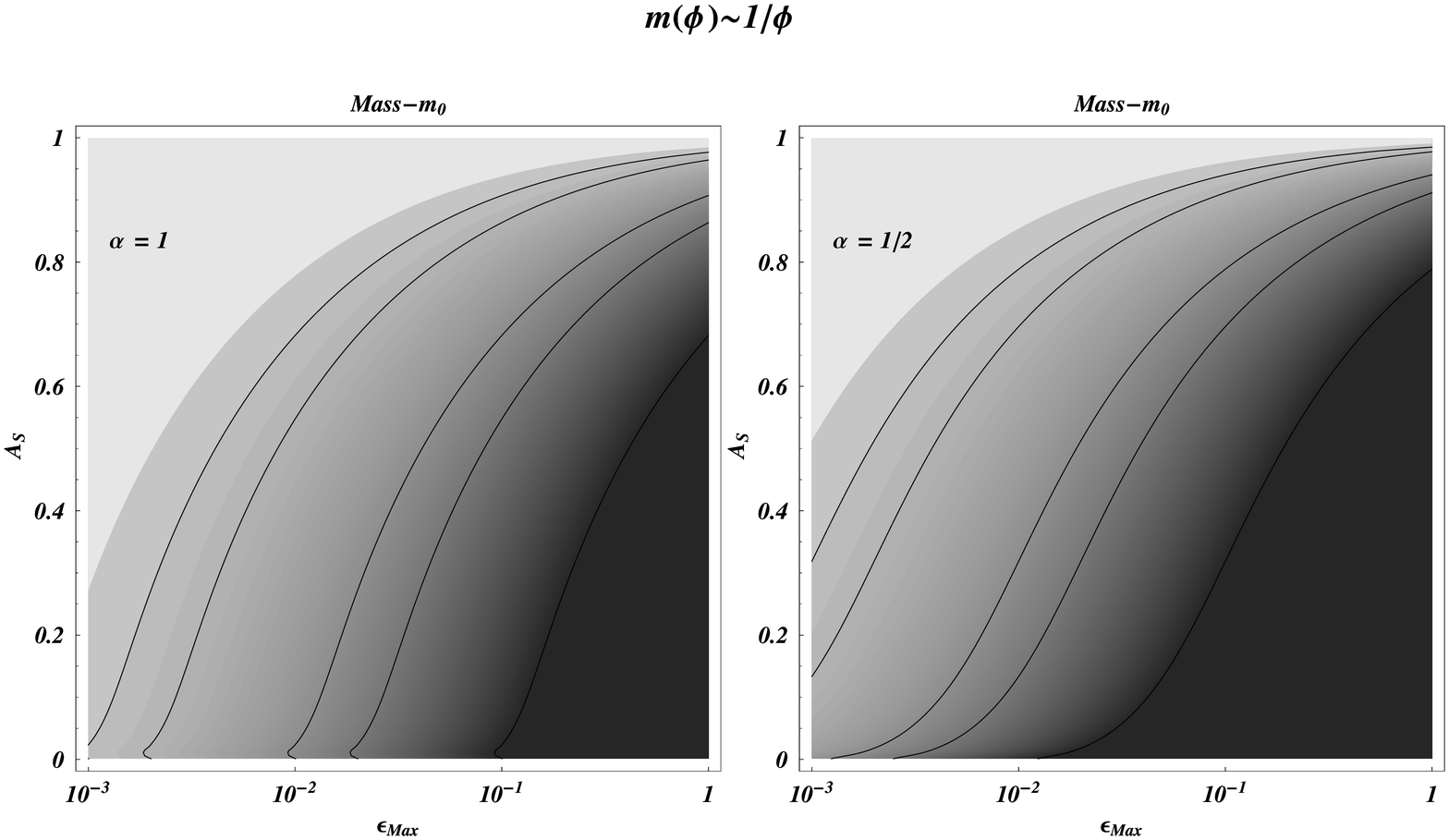,width=13 cm}}
\vspace{-0.5 cm}
\centerline{\psfig{file=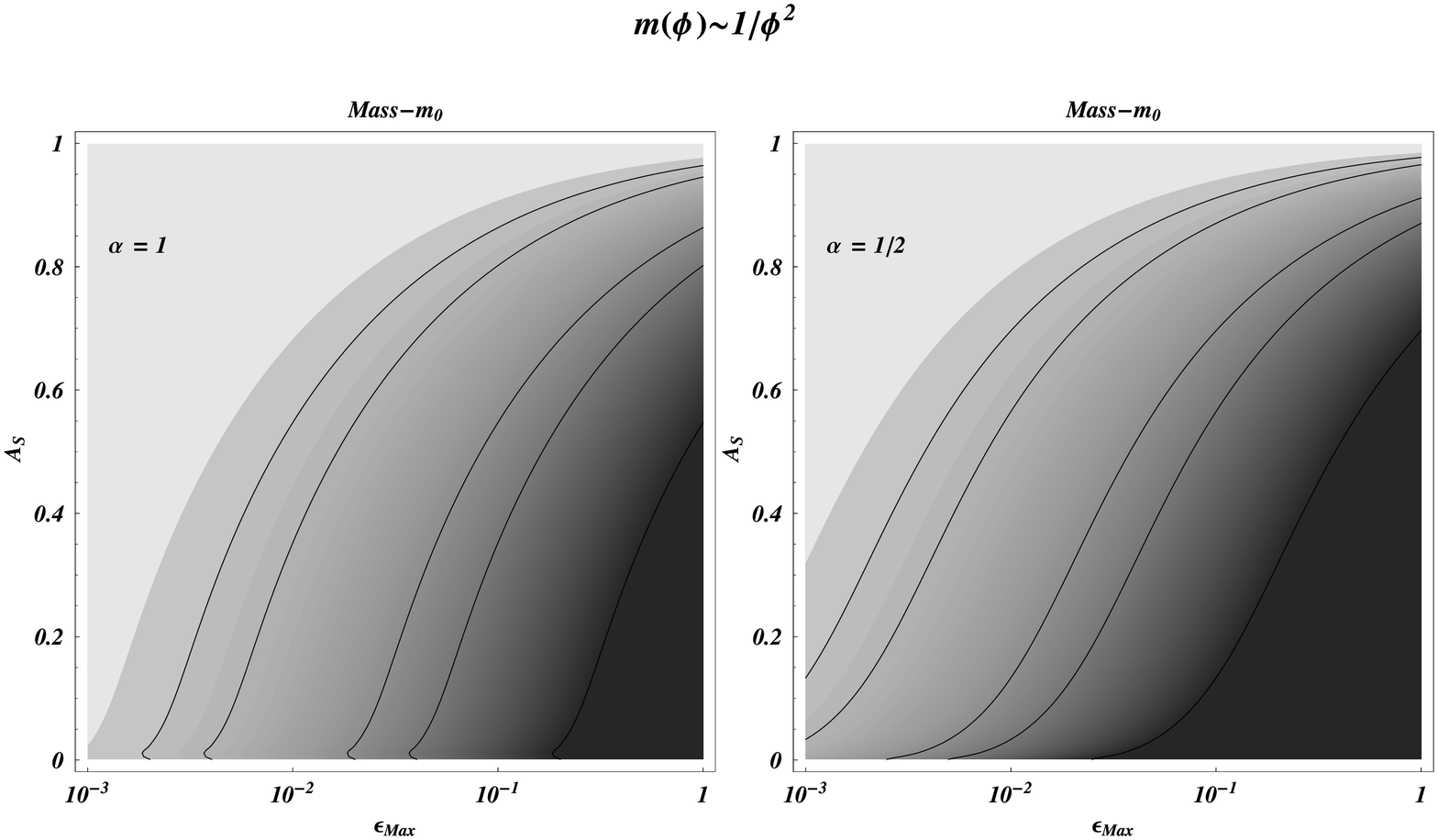,width=13 cm}}
\vspace{-1.0 cm}
\caption{\small Maximal value of the linear perturbation coefficient $\epsilon$ as a function of the present-day neutrino mass and of the energy density coefficient $A_{\s}$ for the GCG.
Once again we have set the neutrino mass varying as $1/\phi$ and $1/\phi^{\2}$ in a GCG scenario for $\alpha = 1,\ 1/2$.
The increasing {\em gray level} corresponds to increasing values of $m_{\0}$, for which we have marked the boundary values for $m_{\0} = 0.05\, eV,\,0.1\, eV,\,0.5\, eV,\,1\,eV,\, 5\, eV$.}
\label{fGCG-05}
\end{figure}

The consistency of our perturbative approach depends on the maximal value assumed by the $\epsilon$ parameter at present (cf. Eq.~(\ref{gcg13})).
In order to estimate $\epsilon$, we use Fig.~\ref{fGCG-04}.
From Eq.~(\ref{gcg24}) we observe that the dominant analytical dependencies of $\phi$ and $V\bb{\phi}$ on the scale factor can, in good approximation, be expressed as $\phi\bb{a}\simeq \kappa_{\1} \ln{(a)}$ and $V\bb{\phi\bb{a}} \simeq \kappa_{\2} + \kappa_{\3}a^{\mi\3(\al \pl \1)}$ in the interval $0 < a < 1$, where $\kappa_{i}$, $i = 1,\,2,\,3,\,...$ are arbitrary constants.
These lead to a dependence of $\epsilon$ on the scale factor given approximately by
\begin{equation}
|\epsilon| \propto \frac{a^{\mi\3\al \pl \1}}{\ln^{\n + \1}(a)\left(\ln(a) + \kappa_{\4}\right)}.
\label{gcg34}
\end{equation}
This is an increasing function of the scale factor dominated by the mass dependent term contribution, $m_{\nu}\bb{\phi{\bb{a}}} \propto \phi\bb{a}^{\mi\n}$, $n = 1,\,2$.
By noticing that $\epsilon_{\mbox{\tiny Max}} = \epsilon\bb{a = 1}$, we show in Fig.~\ref{fGCG-05} the dependence of $\epsilon_{\mbox{\tiny Max}}$ for which GCG parameters and $m_{\0}$ values are consistent with our perturbative approach.

In the next section we analyze the phenomenologically interesting set of parameters for our GCG-neutrino coupled model. As mentioned, the choice of $A_{\s} = 0.7$ will be considered as it is consistent with all known data.

\section{The stability condition as a window for the neutrino phenomenology}

Current cosmological data constrain the number of active neutrino flavours as well as the sum of their masses to $\displaystyle\sum_{i = \1}^{\3} m_{\nu i}\,  <\,  0.75 \,eV$ at $95\,\%$  c.l. \cite{Han06}.
This constraint does not agree with the Heidelberg-Moscow bounds arising from of neutrinoless double beta decay which sets the limit $\displaystyle\sum_{i = \1}^{\3} m_{\nu i}\,  >\,  1.2 \,eV$ at $95\,\%$  c.l. \cite{Kla04}, which, however, are in agreement with the CMB data analysis of the WMAP results that sets $0.7 \,eV$  at $95\,\%$ c. l. for each neutrino species (and $2.0\, eV$ in total) \cite{Ich06}.
Actually, improvement on experimental data are expected to be sensitive to the effects of a finite sum of neutrino masses as small as $0.06\, eV$ \cite{Han06,Pas07}, the lower limit arising by neutrino oscillation experiments that set $\Delta m^{\2} \sim  7.5$ - $8.7 \times 10^{\mi\5} \, eV^{\2}$ ($2 \sigma$) for solar neutrinos, and $\Delta m^{\2} \sim  1.7$ - $2.9 \times 10^{\mi\3} \, eV^{\2}$ ($2 \sigma$) for atmospheric neutrinos.
In addition, current neutrino direct mass measurements, for instance, through the tritium beta decay Mainz experiment, does set an upper limit on the effective electron neutrino mass of $m_{\nu}\,  <\,  2.3 \,eV$ at $95\%$ c.l. \cite{Kra05}.
Thus, in order to test our scenario, we shall consider present neutrino masses varying from $0.05 \, eV$ to $5 \, eV$.

On the other hand, studies of the GCG parameters using supernova and CMB data as well as cosmic topology \cite{Ber03,Ber04,Ber04,Ber06} allow choosing a typical value $A_{\s} = 0.7$, for about $70\%$ of dark energy and $25\%$ of dark matter in a universe filled with $95\%$ by the GCG fluid.

\begin{figure}
\centerline{\psfig{file=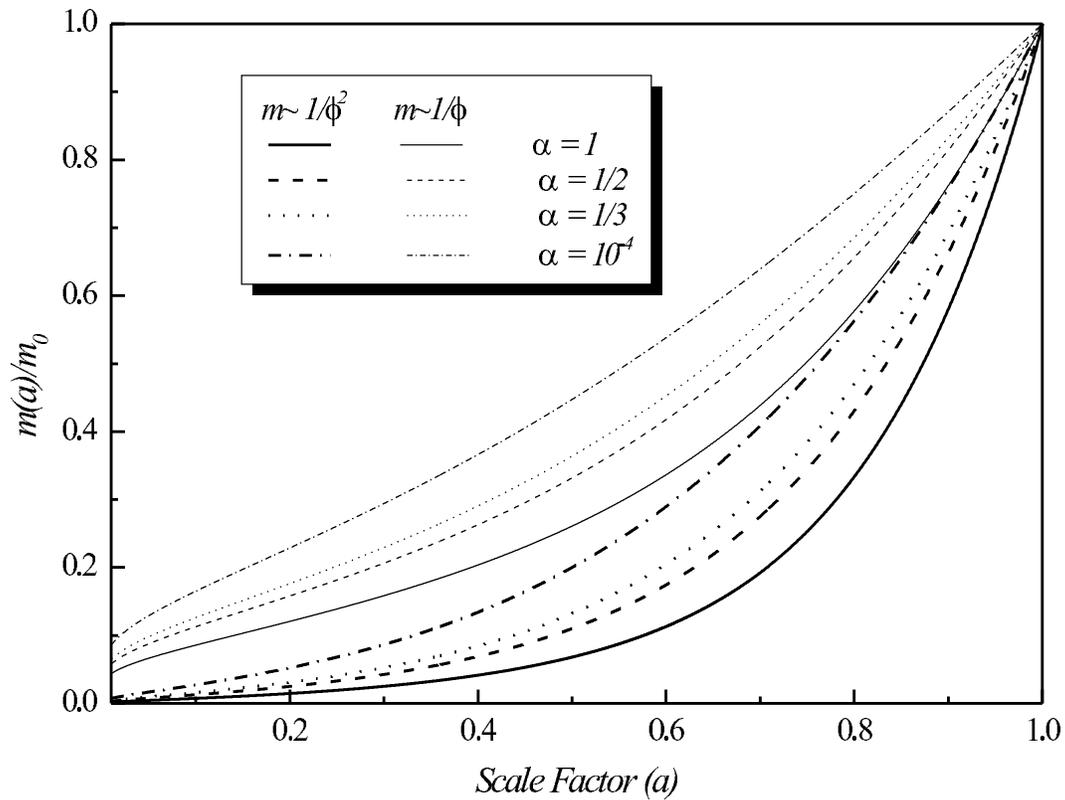,width=14 cm}}
\caption{\small
The neutrino mass ($m_{\nu}\bb{a}/m_{\0}$) dependence on the scale factor $a$.
We have set model dependent neutrino masses varying as $1/\phi$ and $1/\phi^{\2}$ in a GCG scenario with $A_{\s}= 0.7$ and $\alpha = 1,\,1/2,\,1/3,\, 10^{\mi\4}$.}
\label{fGCG-06}
\end{figure}

With these values we can illustrate the behaviour of the neutrino mass in terms of the GCG parameters as a function the scale factor.
This is depicted in Fig.~\ref{fGCG-06}.
One sees that for $m_{\nu}\sim 1/\phi$ neutrinos become NR earlier than the $m_{\nu}\sim 1/\phi^{\2}$ case.
Actually, the stronger is the inverse power dependence of the neutrino mass on the scalar field, that is $m_{\nu}\sim \phi^{\mi\n}$, $n > 1$, the later neutrinos become NR.
If on the other hand, the mass generation model were exponential type, $m_{\nu} \sim \exp{-[\phi/\phi_{\0}]}$, then the neutrinos would become NR much closer to present.

\begin{figure}
\centerline{\psfig{file=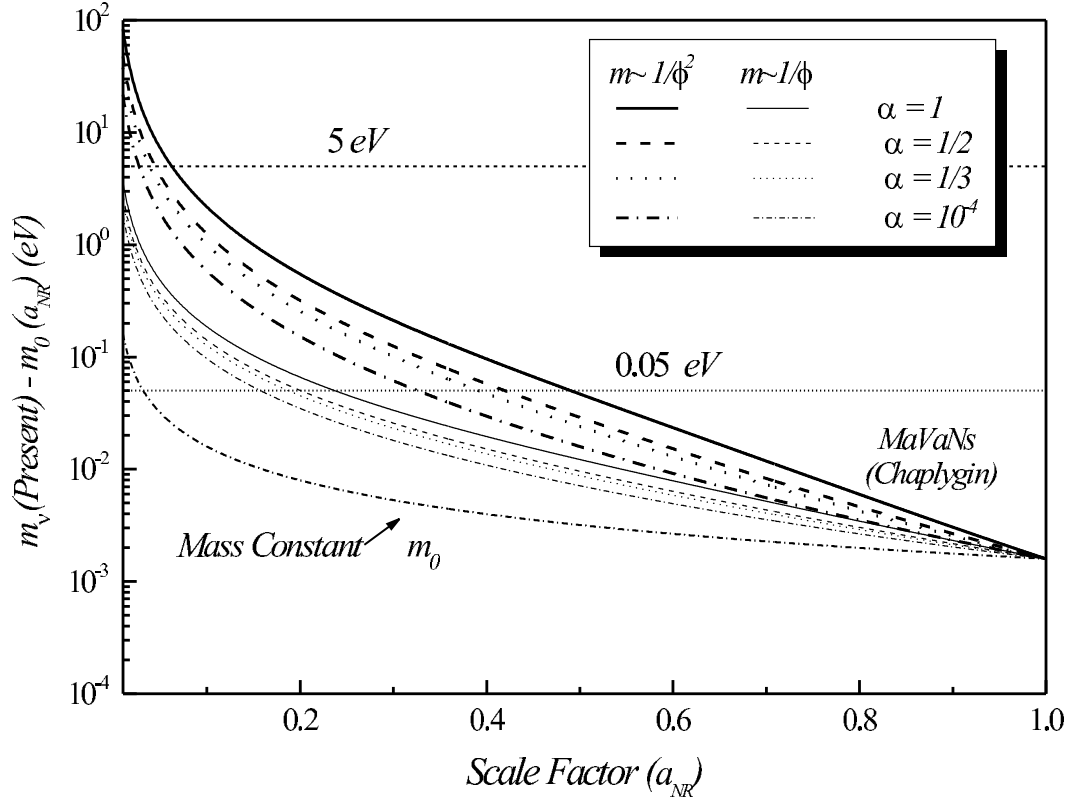,width=14 cm}}
\caption{\small Present-day values of the neutrino mass $m_{\0}$ and the corresponding values of $a_{\N\R}$ for which the transition between the NR and UR regimes takes place in a GCG phenomenological scenario with
$A_{\s} = 0.7$ and $\alpha = 1,\,1/2,\,1/3,\, 10^{\mi\4}$.
The choice of the model for mass generation plays a relevant role in determining the starting point $a_{\N\R}$ of the coupling effectiveness.
This is a section of graphs of Fig.~\ref{fGCG-03} for $A_{\s}= 0.7$.}
\label{fGCG-07}
\end{figure}

\begin{figure}
\centerline{\psfig{file=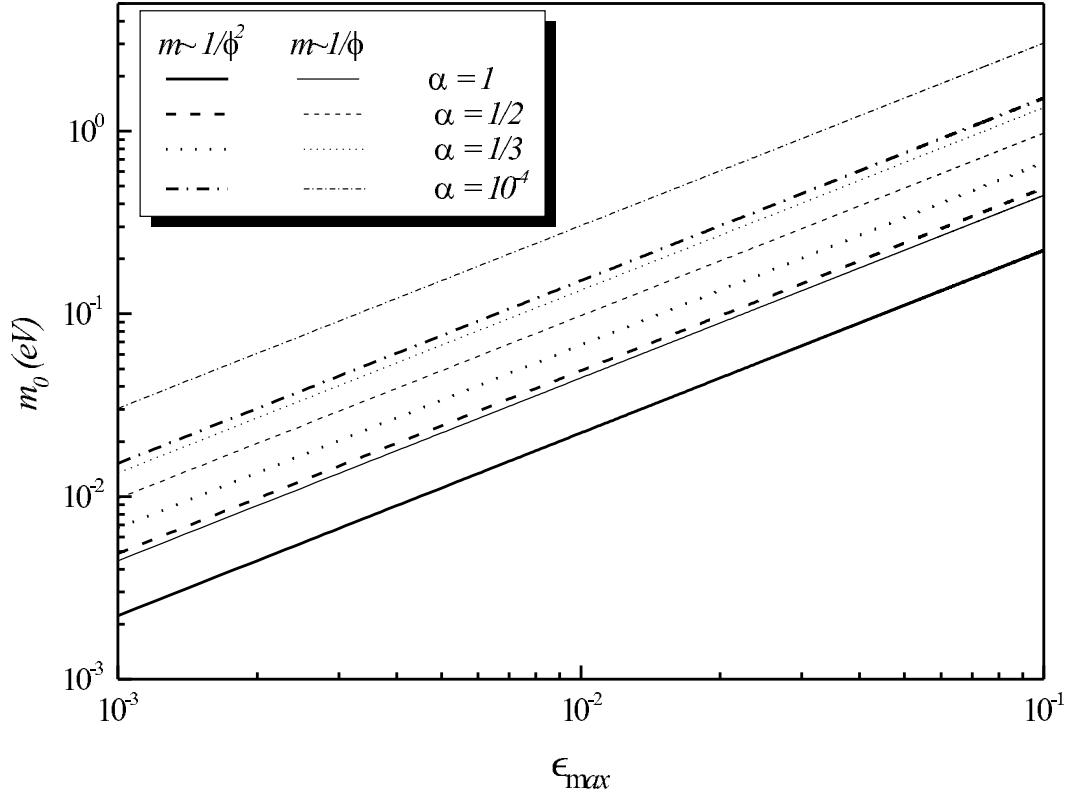,width=14 cm}}
\caption{\small The present-day neutrino mass $m_{\0}$ with respect to the maximal value of the linear perturbation coefficient $\epsilon$ for $A_{\s} = 0.7$ in a GCG scenario with $\alpha = 1,\,1/2,\,1/3,\, 10^{\mi\4}$.
As before, the choice of the model for mass generation plays a relevant role in determining the range of validity of the perturbative approach.
Notice that the maximal value for $\epsilon$ corresponds to $a = 1$.}
\label{fGCG-08}
\end{figure}

In addition, from the constraint set by Eq.~(\ref{gcg33}), which establishes the neutrino NR regime, we show in Fig.~\ref{fGCG-07}, for the same set of parameters, present-day values of the neutrino masses and the corresponding values of $a_{\N\R}$ for which the transition between the NR and UR regimes takes place.
Since there are strong phenomenological constraints on the choice of $m_{\0}$, it is important to pay attention to the present-day value of neutrino mass interval, from $0.05\, eV$ to $5\, eV$, where a clear dependence on the model for mass generation is observed.
The $a_{\N\R}$ values found in Fig.~\ref{fGCG-07} are taken into account in Fig.~\ref{fGCG-08}, where it is examined the validity of the perturbative approach for different GCG scenarios.

\begin{figure}
\centerline{\psfig{file=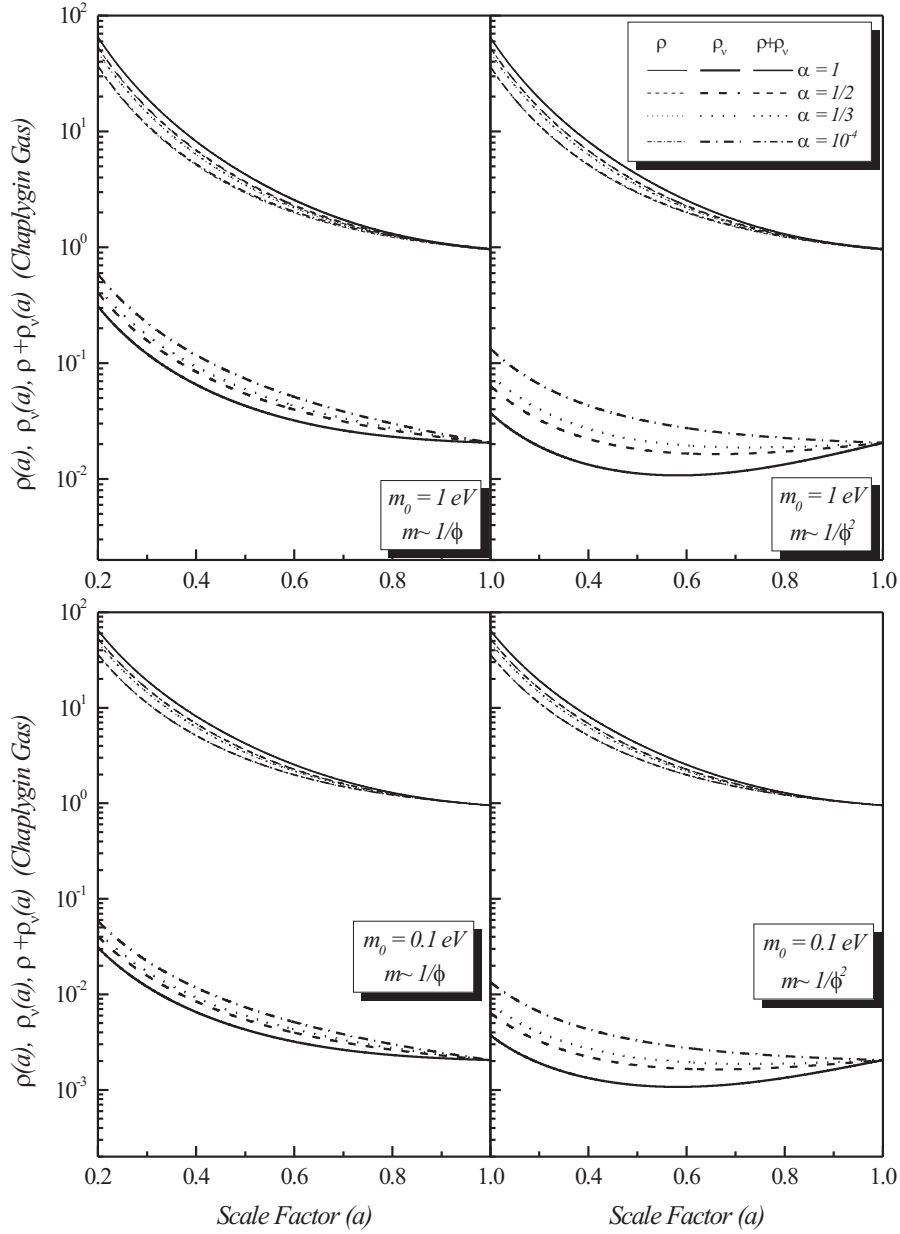,width=14 cm}}
\caption{\small Components of the energy density (in units of $\rho_{\mbox{\tiny Crit}}$) obtained from our perturbative approach for $A_{\s} = 0.7$ in a GCG scenario with $\alpha = 1,\, 1/2,\,1/3,\, 10^{\mi\4}$.
The choice of the model for mass generation does not play a relevant role on behaviour of the modified energy components.}
\label{fGCG-09}
\end{figure}

By observing the model dependent conditions discussed above, namely the value of $a = a_{\N\R}$, and the present-day neutrino mass corresponding to the maximal value of the linear perturbation coefficient $\epsilon_{\mbox{\tiny Max}}$, we can, for instance, set the phenomenologically acceptable values of $m_{\0} = 1\, eV$ and $m_{\0} = 0.1 \, eV$, in order to perturbatively quantify the modifications to the energy density components of the coupled fluid.
Such perturbative corrections are shown in Fig.~\ref{fGCG-09}.

Interestingly, for $m_{\0} = 1\, eV$, a fairly typical value, we can see in Fig.~\ref{fGCG-10} that stable MaVaN perturbations correspond, for the GCG case, to a well defined effective squared speed of sound speed
\begin{equation}
c_{s}^{\2} \simeq \frac{d p_{\phi}}{d(\rho_{\phi}+ \rho_{\nu})} > 0.
\end{equation}
The larger are the $m_{\0}$ values, the larger are the corrections to the squared speed of sound, up to the limit where the perturbative approach cannot be applied.
Therefore, as far as the perturbative approach is valid, our model does not run into stability problems in the NR neutrino regime.
In opposition, in the usual treatment where neutrinos are just coupled to dark energy, cosmic expansion together with the gravitational drag due to cold dark matter have a major impact on the stability of MaVaN models.
Usually, for a general fluid for which we know the equation of state, the dominant effect on $c_{s}^{\2}$ arises from the dark sector component and not by the neutrino component.
For the models where the stationary condition (cf. Eq.~(\ref{gcg07})) implies a cosmological constant type equation of state, $ p_{\phi} = - \rho_{\phi}$, one obtains $c_{s}^{\2} = -1$ from the very start of the analysis.
For sure the situation cannot be fixed by the perturbative contribution of neutrinos.
Our GCG-neutrino model is free from this inconsistency.

\section{Conclusions}

In this work we have analyzed the coupling of neutrino masses to a light scalar field associated with the GCG unification model of dark energy and dark matter.
We have presented the criteria of applicability of a perturbative approach for in the study of MaVaN models and determined the coefficient of a linear perturbation which is given in terms of the ratio between the variation of the neutrino energy and scalar field potential terms (cf. Eq.~(\ref{gcg13})).
Seesaw proposals for the neutrino mass were then considered and the effects of the neutrino mass coupling with the underlying scalar field of the GCG model were quantified.
As discussed, our approach yields a positive squared speed of sound and is consistent with the current neutrino mass experimental limits.

For sure, adopting a perturbative approach is equivalent to the assertion that the coupling between neutrinos and dark sector is fairly weak.
Besides, our proposal turns out to be an interesting alternative to the usual stationary condition constraint proposed in the Ref. \cite{Far04} for the equation of state $p_{\phi}= - \rho_{\phi}$.
For several configurations, the latter scenario leads to catastrophic instabilities associated with an imaginary speed of sound in the neutrino NR regime.
Actually, in previous work it has been pointed out that MaVaN models generically face stability problems for some choices of neutrino-scalar field couplings and scalar field potentials once neutrinos become NR.
Effectively, the scalar field mediates an attractive force between neutrinos which can lead to the formation of neutrino nuggets.
This would turn the combined fluid to behave like cold dark matter and thus render it non-viable as a candidate for dark energy.

\begin{figure}
\centerline{\psfig{file=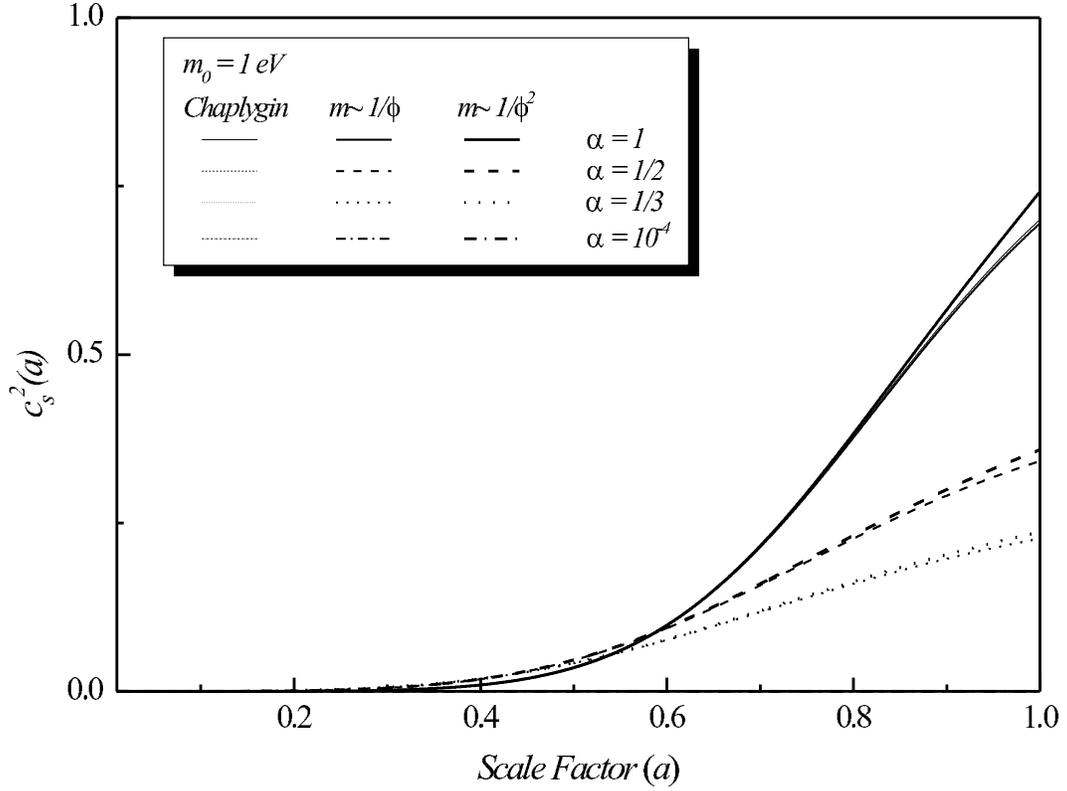,width=14 cm}}
\caption{\small Perturbative modification on the square of the speed of sound $c^{\2}_{s}\bb{a}$ as a function of the scale factor for the neutrino-GCG coupled fluid in comparison with the adiabatic GCG fluid for $A_{s} = 0.7$.
The coupling with the GCG turn the MaVaN's naturally stable.}
\label{fGCG-10}
\end{figure}

In opposition, our analysis shows that the coupling of neutrinos with the scalar field of the GCG model is consistent in what concerns the positiveness of the squared speed of sound (cf. Fig.~\ref{fGCG-10}).
The knowledge of the background fluid equation of state and our perturbative approach allows one to overcome the problematic negative squared speed of sound.
Even though our analysis could be extended to other equations of state, the GCG model seems to be a quite suitable candidate for constructing stable MaVaN scenarios.

To conclude, we stress that allowing for dynamical behaviour to a scalar field associated to dark energy in connection with the SM neutrinos and the electroweak interactions may bring important insights on the physics beyond the SM.
Neutrino cosmology, in particular, is a fascinating example where salient questions concerning SM particle phenomenology can be addressed and hopefully better understood.
We believe that our proposal is a further step in this respect.

\begin{acknowledgments}
A. E. B. would like to thank the financial support from the FAPESP (Brazilian Agency) grant 07/53108-2 and the hospitality of the Physics Department of the Instituto Superior T\'{e}cnico, Lisboa, Portugal, where this work was carried out.
O. B. would like to acknowledge the partial support of Funda\c{c}\~ao para Ci\^encia e Tecnologia (Portuguese Agency)
under the project POCI/FIS/56093/2004.
\end{acknowledgments}

\pagebreak
\newpage

\end{document}